\documentclass[acmsmall]{acmart}
\AtBeginDocument{%
  \providecommand\BibTeX{{%
    \normalfont B\kern-0.5em{\scshape i\kern-0.25em b}\kern-0.8em\TeX}}}

\setcopyright{acmlicensed}
\acmJournal{PACMCGIT}
\acmYear{2021} \acmVolume{4} \acmNumber{3} \acmArticle{} \acmMonth{9} \acmPrice{15.00}\acmDOI{10.1145/3480140}

\acmSubmissionID{14}

\usepackage{amscd} 
\usepackage{amsfonts} 
\usepackage{amsmath} 
\usepackage{hyperref} 
\usepackage{subfigure} 

\usepackage[utf8x]{inputenc}

\usepackage{placeins}
\usepackage[algo2e]{algorithm2e} 

\SetAlFnt{\small}
\SetAlCapFnt{\small}
\SetAlCapNameFnt{\small}
\SetAlCapHSkip{0pt}

\usepackage{algorithm}
\usepackage[noend]{algpseudocode}
\usepackage{array}
\usepackage{overpic}
\usepackage{bm}
\usepackage{wrapfig}
\usepackage{pgfplots}
\usepackage{pgf,tikz}
\usepackage{soul}
\usepackage[normalem]{ulem}

\newcommand{\M}{\mathcal{M}}
\newcommand{\N}{\mathcal{N}}

\begin{document}

\title{A functional skeleton transfer}

\author{Pietro Musoni}
\affiliation{%
  \institution{University of Verona}
  \city{Verona}
  \country{Italy}
}
\email{pietro.musoni@univr.it}

\author{Riccardo Marin}
\affiliation{%
  \institution{Sapienza University of Rome}
  \city{Rome}
  \country{Italy}
}
\email{marin@di.uniroma1.it}

\author{Simone Melzi}
\affiliation{%
  \institution{Sapienza University of Rome}
  \country{Italy}
  }
\email{melzi@di.uniroma1.it}

\author{Umberto Castellani}
\affiliation{
  \institution{University of Verona}
  \city{Verona}
  \country{Italy}
}
\email{umberto.castellani@univr.it}

\renewcommand{\shortauthors}{Musoni, et al.}






\begin{abstract}
The animation community has spent significant effort trying to ease rigging procedures. This is necessitated because the increasing availability of 3$D$ data makes manual rigging infeasible. However, object animations involve understanding elaborate geometry and dynamics, and such knowledge is hard to infuse even with modern data-driven techniques. Automatic rigging methods do not provide adequate control and cannot generalize in the presence of unseen artifacts. As an alternative, one can design a system for one shape and then transfer it to other objects. In previous work, this has been implemented by solving the dense point-to-point correspondence problem. Such an approach requires a significant amount of supervision, often placing hundreds of landmarks by hand.
This paper proposes a functional approach for \emph{skeleton transfer} that uses limited information and does not require a complete match between the geometries. To do so, we suggest a novel representation for the skeleton properties, namely the \emph{functional regressor}, which is compact and invariant to different discretizations and poses. We consider our functional regressor a new operator to adopt in intrinsic geometry pipelines for encoding the pose information, paving the way for several new applications.
We numerically stress our method on a large set of different shapes and object classes, providing qualitative and numerical evaluations of precision and computational efficiency. 
Finally, we show a preliminar transfer of the complete rigging scheme,  introducing a promising direction for future explorations.
\end{abstract}



\begin{teaserfigure}
  \includegraphics[width=\textwidth]{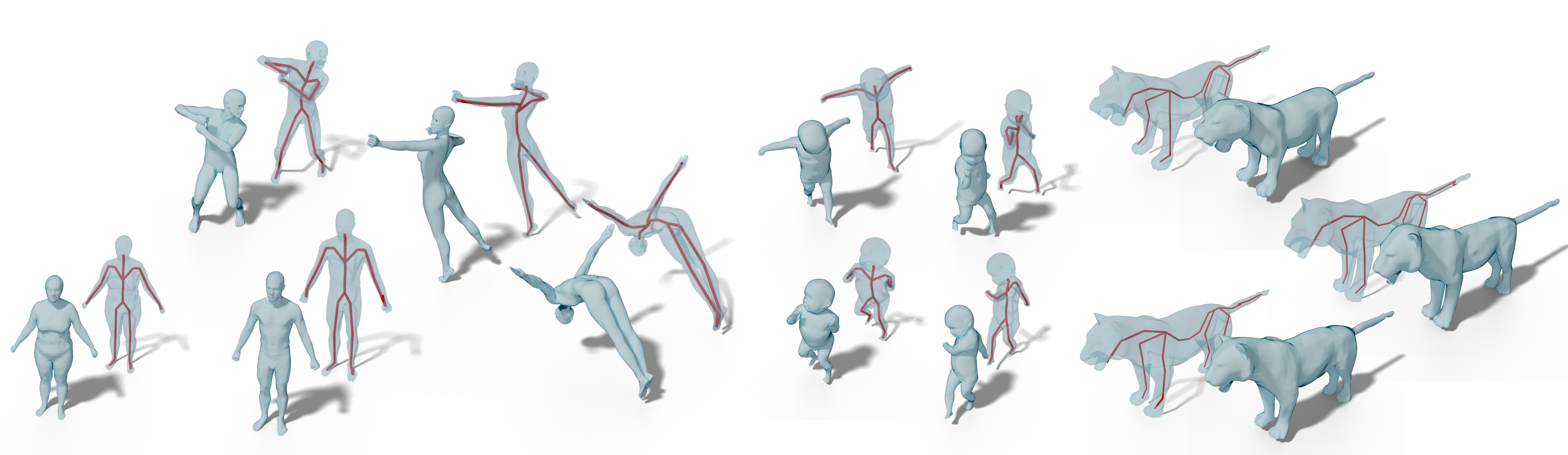}
  \caption{Some results of our method. On the left, five adult humans that inherited the SMPL \cite{loper2015smpl} skeleton; these surfaces come from different datasets and have different discretizations. In the middle, four children shapes on which we transfer the skeleton from SMIL \cite{Hesse2018}. On the right, three lions that received the skeleton from SMAL \cite{smal}.}
  \label{fig:teaser}
\end{teaserfigure}

\maketitle

\section{Introduction}
\label{sec:introdution}
Animating a 3D object is a time-consuming and challenging problem; it requires matching the expectations suggested by our experience and balancing the trade-off between usability and expressiveness. Also, a coherent rigging system across similar objects is desirable to allow reuse of animations with little effort. 

Several methods have been proposed to solve the automatic rigging of meshes. They rely on algorithms that process the geometric structure \cite{Jacobson2012} or, more recently, by learning using data-driven approaches \cite{rajendran2020virtual}. 
However, they hardly produce coherent results among arbitrary objects and do not provide the user with additional information about the semantics and the structural role of each component of the shape.
%
To keep a degree of control, another approach is defining a rigging system and transferring it to other models, but this is commonly achieved by establishing a dense point-to-point correspondence, which requires user annotations. Also, determing point-to-point correspondence is one of the fundamental open problems of geometry processing, particularly when facing non-isometric objects. 

In this work, we claim that the skeleton can be accurately transferred by a functional approach without relying on a dense point-to-point correspondence. Our pipeline asks the user to define a rigged system for one shape. Then, we optimize for a regressor that is parsed to a functional representation, and finally, it is transferred to other shapes. This representation offers a new perspective in skeleton encoding; it is compact and entirely intrinsic (i.e., it does not change with varying poses of the subject).
Given a new shape to animate, we ask for a few landmarks on the target object (seven on humans and which, in practice, can be detected almost automatically) and we rely on the Functional Maps framework \cite{nogneng2017informative} to transfer our rigging. This framework is state-of-the-art for the functional transferring task, both in terms of performance and efficiency. Our approach produces results coherent across different objects of the same class, as shown in Figure \ref{fig:teaser}. Finally, we cannot find implementations for previous methods; we consider this one of the main reasons for proposing novel approaches. The code and data used in this paper will be made entirely available for research purposes.

To summarize, our contribution is threefold:
\begin{enumerate}
\item We propose the first functional pipeline for skeleton transfer. It produces accurate results efficiently and with limited user input. It is general across different settings and object classes.
\item We introduce a new spectral representation for the shape skeleton information; that is compact, entirely intrinsic, and invariant to different discretizations and poses of the subject. This representation provides a new perspective to analyze object mechanics.
\item We will release all the code and data used in our experiments.
\end{enumerate}

\section{State of the art}
\label{sec:stateofart}

\subsection{Classical methods}
Several researchers have tried to define the skeleton for a given shape by using automatic systems. One of the fundamental works in this field is Pinocchio \cite{baran2007automatic} by Baran and Popovic; the proposed pipeline produces an animation-ready character starting from a generic object. This process is divided into two steps: skeleton embedding and skin attachment. 

Some works \cite{gleicher1998retargetting}, \cite{lee1999hierarchical}, \cite{choi2000online} solve kinematic constraints on joint positions using displacement maps.
In \cite{allen2003space}, the authors provide joint retargeting through the use of three markers. The distance between the joints and the markers should be fixed. However, this method works only on near-similar morphological bodies.
Methods like \cite{capell2005physically} and \cite{orvalho2008transferring} introduced the transferring of the rig. The first is not compatible with common skinning methods, and the second can only be applied to a limited set of characters.
In \cite{seo2010rigging}, the authors provide a method based on data mapping applicable to surfaces and volumes, but a careful manual selection of points is necessary to initialize the correspondence. 

\subsection{Automatic Rigging and rigging transfer}
The main disadvantage of automatic rigging methods is that the animation sequence should be designed on a specific skeleton. Instead, the rigging transfer methods tackled this problem by transferring a native skeleton with an attached animation, combining it to the new shape.
Several works focus on the transfer of the medial-axis \cite{yang2018dmat} \cite{seylan20193d}, which is the set of points having more than one closest point to the boundary of a shape. The medial axis of a 3D object can be expressed both as a curve and as a 2D surface. However, medial axis methods are not compatible with standard render engines, and they are computationally inefficient.

Another approach explored in \cite{smooth2012le} and \cite{robust2014le} employs example-based methods to automatically generate linear blend skinning models. The method takes as input a sequence of poses of a model and it outputs the skeleton, the skinning weights and the bone transformations. The output is compatible with common game engines and 3D modelling software, however, such a sequence of shapes is not always easy to have since requires data acquisition from a real-world scenario and the method largely depends on the quality of input example meshes.   

\subsection{Motion transfer}
Other approaches \cite{liu2018surface}, \cite{basset2019contact} propose to transfer each pose of a 3D shape to another using a non-rigid registration. In these methods, a further adjustment is required to have a smooth animation from the sequence of poses obtained.
A more promising procedure involves deforming a shape to adapt to a rigged one, applying the given skeleton to the deformed shape.
 In \cite{marin2020farm}, \cite{CMH} and \cite{ni2020scan}, the authors made a non-rigid registration from a source shape to a target one, and then transfer the skinning information.
 A similar work \cite{jin2018aura} introduces the idea of aura mesh, a volume that surrounds the character and permits avoidance of self-collisions.
 The rigless methods \cite{basset2019contact} do not need the post-processing angle correction, but they require additional steps to obtain an animation-ready shape as output.
In \cite{avril2016animation} the authors compute a pointwise matching between the two shapes to transfer all the properties of an animation setup. This method requires the manual input of more than 300 key points to initialize the dense correspondence. Then, a regression matrix is estimated to compute the joint positions of the new shape. In addition to the high amount of user input, the method requests the two shapes to be in the same pose.
A joint-regressor was also introduced for the SMPL morphable model \cite{loper2015smpl}, but it requires learning from a large dataset of 3D shapes. Furthermore, the joint-regressor of SMPL is defined pointwise, hence its transfer would require a dense point-to-point correspondence.

\section{Background}
\label{sec:background}

\subsection{LBO basis on shape}
\begin{figure}[!t]
 \begin{center}
  \begin{overpic}
  [trim=0cm 0cm 0cm 0cm,clip, width=0.99\linewidth]{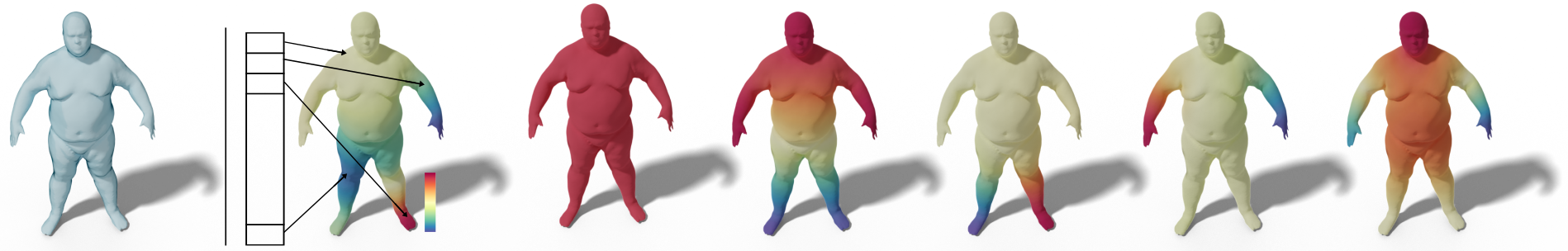}
  \put(5,-2){\scriptsize $\M$}
  \put(25,-2){\scriptsize  $f$}
  \put(38,-2){\scriptsize  $\Phi_0$}
  \put(52,-2){\scriptsize  $\Phi_1$}
  \put(65, -2){\scriptsize  $\Phi_2$}
  \put(78, -2){\scriptsize  $\Phi_3$}
  \put(90, -2){\scriptsize  $\Phi_4$}
  
  \put(31,8){\scriptsize  =}
  \put(43.5,8){\scriptsize  $\alpha_1 +$}
  \put(56.5,8){\scriptsize  $\alpha_2 +$}
  \put(69.5,8){\scriptsize  $\alpha_3 +$}
  \put(82.5,8){\scriptsize  $\alpha_4 +$}
  
  \put(97,8){\scriptsize  \dots}
  \put(16.5,12.8){\tiny  0}
  \put(15.7,11.5){\tiny  -0.2}
  \put(16,10.2){\tiny  1.2}
  \put(16.3,7){\tiny  \dots}
  \put(15.7,0.5){\tiny  -1.4}
  
  \put(27.8,5) {\tiny max}
  \put(27.8,1) {\tiny min}
  \end{overpic}
\end{center}
\caption{\label{fig:functional} Here we visualize some of the functional quantities we referred into the text. Given a geometry $\M$, a function $f$ is a vector that associates a scalar value to each point of the surface. Such function can be decomposed for some basis $\mathbf{\Phi}$ (e.g. the LBO eigenfunctions) recovering a list of coefficients $\alpha$.}
\end{figure}
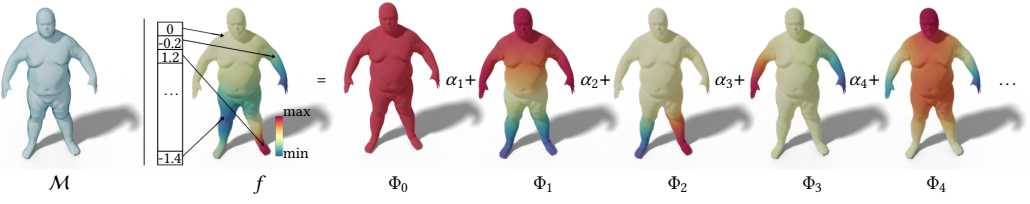
We describe $2$-dimensional surfaces embedded in $\mathbb{R}^3$ as compact and connected 2-dimensional Riemannian manifolds. Given a surface $\M$, we consider $\mathcal{F}(\M):= \lbrace f: \M \rightarrow \mathbb{R}\rbrace$ as the space of real-valued functions defined over it. In other words, a function $f\in \mathcal{F}(\M)$ associates at every point $x\in \M$ a real value  $ f(x) \in \mathbb{R}$. 
In the discrete setting, we encode the shape $\M$ in a triangular mesh, defined by the 3D coordinates of its $n_{\M}$ vertices $X_{\M}\in\mathbb{R}^{n_{M}\times 3}$ and the list of the edges $E_{\M}$ that connect these vertices generating triangular faces.
We associate to each mesh the cotangent Laplace-Beltrami operator (LBO), that is discretized by a $n_{\M}\times n_{\M}$ matrix~\cite{meyer03}. We collect the first $k$ eigenfunctions of the LBO of $\M$ (the ones associated to the $k$ smallest eigenvalues) as columns in a matrix $\bm{\Phi}_{k_{\M}} = \big[\phi_1, \phi_2,\cdots, \phi_{k_{\M}}\big]$. In geometry processing, these eigenfunctions are widely used to approximate functions or signals defined on 3D shapes extending Fourier analysis on non-Euclidean domains~\cite{Taubin,Levy06,Levy08}.
In particular, functions can be represented by a linear combination of the basis  $\bm{\Phi}_{k_{\M}}$ as $\tilde{f} = \bm{\Phi}_{k_{\M}}\bm{\widehat{f}}$, where $\bm{\widehat{f}}$ are the basis coefficients. Since just a subset of the eigenfunctions is used, this is just an approximation of $f$. The coefficients $\bm{\widehat{f}}$ can be obtained by the projection $\bm{\widehat{f}}=\bm{\Phi}_{k_{\M}}^\dagger f$, where $\bm{\Phi}_{k_{\M}}^\dagger$ is the Moore–Penrose pseudo-inverse (i.e. $\bm{\Phi}_{k_{\M}}^\dagger\bm{\Phi}_{k_{\M}} = Id_{k_{\M}\times k_{\M}}$).
The matrix $\bm{\Phi}_{k_{\M}}^\dagger$ could be seen as the operator that projects the pointwise values of the function into its Fourier representation. We will exploit this projection through the paper, and we refer to this space of the Fourier coefficients as the functional representation of quantities defined pointwise on the mesh. A visualization of these quantities is shown in Figure \ref{fig:functional}.

\subsection{Functional Maps}
\label{sec:fmaps}
A correspondence $T$ between two surfaces $\mathcal{N}$ and $\mathcal{M}$ is a mapping that associates to each point $p\in \N$ the corresponding point $T(p)\in \M$.
In the discrete case, we can encode this mapping in a matrix $\Pi_{\N\M}\in \mathbb{R}^{n_{\M}\times n_{\N}}$ such that $\Pi(i,j) = 1$ if the $j$-th vertex of the shape $\N$ corresponds to the $i$-th vertex of the shape $\M$ and $\Pi(i,j) = 0$ otherwise.
From a functional perspective, any function $f\in\mathcal{F}(\M)$ could be transferred to a function $g = f \circ T \in \mathcal{F}(\N)$. 
Hence, given the truncated bases $\bm{\Phi}_{k_{\M}}$ and $\bm{\Psi}_{k_{\N}}$ obtained from the eigendecomposition of the LBO respectively of $\M$ and $\N$, the Functional Maps framework \cite{ovsjanikov2012functional} exploits these truncated bases to encode this mapping in a compact matrix $\bm{C} \in \mathbb{R}^{k_{\N}\times k_{\M}}$ referred as a \emph{functional map}.
The matrix $\bm{C}$ can be directly derived from $\Pi$ with the following equation $\bm{C} = \bm{\Psi}_{k_{\N}}^\dagger\Pi_{\N\M}\bm{\Phi}_{k_{\M}}$.
Once we have a functional map $\bm{C}$, we can recover the correspondence $T$ by transfering an indicator function for each vertex, but \cite{ovsjanikov2012functional} show that this is equivalent to the nearest-neighbour assignment problem between $\bm{\Phi}_{k_{\M}}\bm{C}^{\top}$ and $\bm{\Psi}_{k_{\N}}$.
Different strategies and constraints have been proposed to solve for the matrix $\bm{C}$ \cite{ovsjanikov2012functional,nogneng2017informative,nogneng18,ren2019structured,ren2018continuous}. Several attempts have also been implemented to extract and refine the pointwise correspondence from the functional representation \cite{rodola-vmv15,ezuz2017deblurring,zoomout,MapTree,pai2021fast}. 
They produce meaningful results, but the conversion to pointwise maps induces distortions even for accurate functional maps.
Here, we adopt the method proposed in \cite{nogneng2017informative} following the procedure described in \cite{zoomout}.

\section{Method}
\label{sec:method}
\begin{figure}[!t]
 \begin{center}
  \begin{overpic}
  [trim=0cm 0cm 0cm 0cm,clip, width=0.99\linewidth]{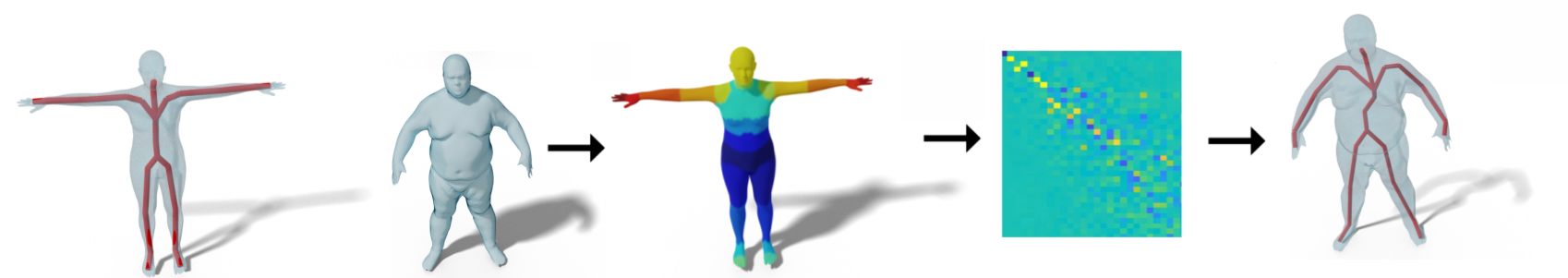}
  \put(8,15){Input}
  \put(26,15){Target}
  \put(41,18){Functional}
  \put(41,15){Optimization}
  \put(64,18){Functional}
  \put(66,15){Map}
  \put(78,18){Skeleton Transfer}
  \end{overpic}
\end{center}
\caption{\label{fig:pipeline} This figure depicts our pipeline overview. Our method inputs are an initial rigged model and a target mesh. Then, we perform optimization to obtain the functional representation of the mesh skinning, and by computing a functional map between the two, we can transfer the skeleton to the target. The transfer lets us move the target following the semantic of the source.}
\end{figure}
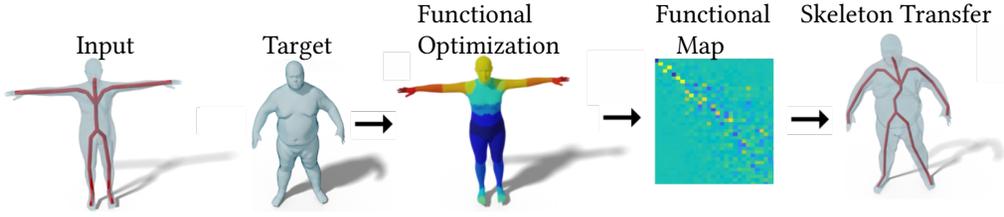

Here we describe our functional-based representation and its relation to the spatial one. Then, we present our algorithm for skeleton transfer, summarized by Figure \ref{fig:pipeline}.

\subsection{Functional Regressor}
A common choice is defining the skeleton $\bm{J}_{\M} \in \mathbb{R}^{Q \times 3}$ as a function of the given surface $\M$, namely the \emph{joint regressor}. Following \cite{loper2015smpl} it can be expressed as a linear operator $\bm{R} \in \mathbb{R}^{Q \times n_{\M}}$, that acts on the coordinates of the vertices $X_{\M}$ in the discrete setting; hence:
\begin{equation}
\begin{aligned}
    \bm{J}_{\mathcal{M}} = \bm{R}X_{\M}.
\label{eq:RegreBaseEq}
\end{aligned}
\end{equation}
We refer to $\bm{R}$ as the standard or spatial regressor.

Starting from this, we propose a novel functional formulation. 
We can consider the coordinates of the vertices of the meshes $X_{\M}$ as a set of three functions (one for each column) and project them in the space of the Fourier coefficients:
\begin{align}
\widehat{X}_{\M} &= \bm{\Phi}_{k_{\M}}^{\dagger} X_{\M}
\end{align}
where the coefficients $\widehat{X}_{\M} \in \mathbb{R}^{k_{\M}\times 3}$.

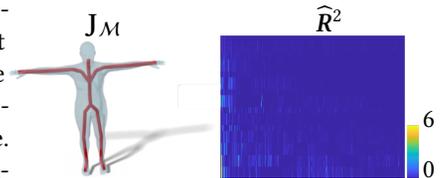
\begin{wrapfigure}[8]{R}{0.4\linewidth}
 \begin{center}
  \hspace{-0.8cm}
 \begin{minipage}{\linewidth}
  \begin{overpic}
  [trim=0cm 0cm 0cm 0cm,clip, width=0.99\linewidth]{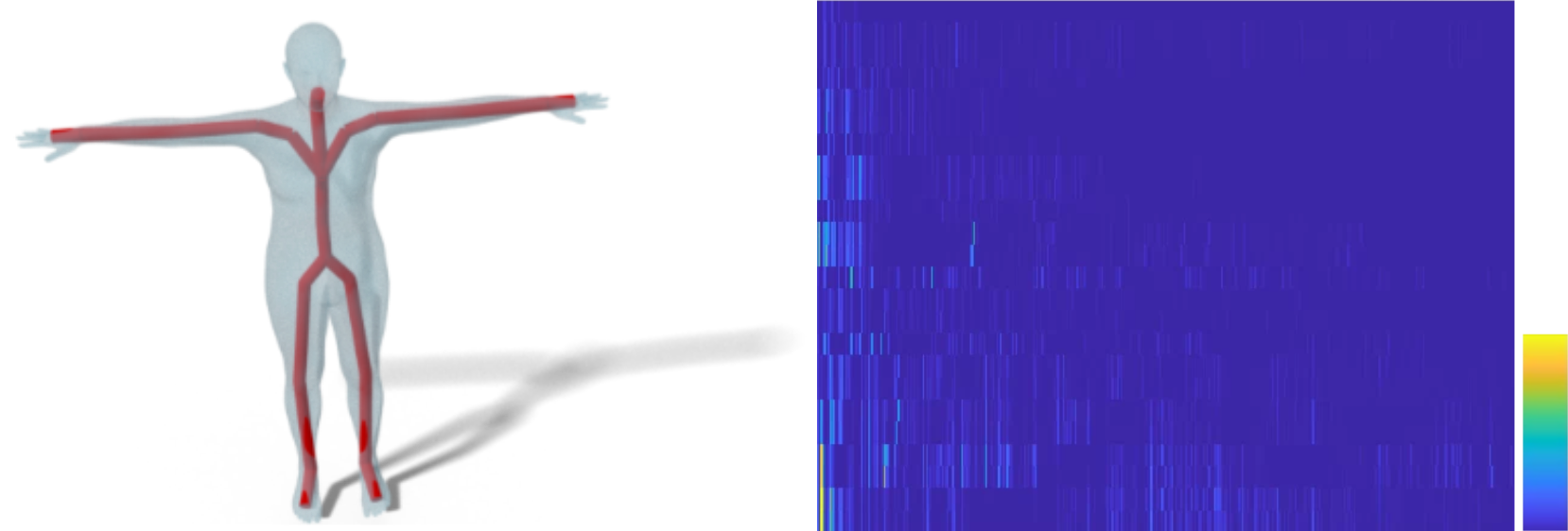}
  \put(101,0){0}
  \put(101,12){6}
  
  \put(20,35){$\mathbf{J_\mathcal{M}}$}
  \put(75,35){$\widehat{\bm{R}}^2$}
  \end{overpic}
  \end{minipage}
\end{center}
\caption{\label{fig:regressor}On the left, the defined skeleton. On the right, the functional regressor.}
\end{wrapfigure}
\vspace{0.5cm}\noindent\textbf{Motivation.} 
Since our functional representation is defined on the low-pass filtering of the 3D coordinates, it captures the global structure of the shape, with resilience to high-frequency details. Also, it is intrinsic; it is agnostic to the subject pose and placement in the 3D space. Moreover, it is compact: a spatial joint regressor is encoded by a matrix $n_{\M} \times Q$; our functional regressor has dimension $k_{\M} \times Q$. We generally select $k_{\M} \in [30, 200]$; smaller than $n_{\M}$ by at least two orders of magnitude.

In this functional representation, we want to find a linear operator $\widehat{\bm{R}}$ such that: 
\begin{align}
\label{eq:specregrX}
\bm{J}_{\mathcal{M}} &= \widehat{\bm{R}} \widehat{X}_{\M}
\end{align}
where we refer to $\widehat{\bm{R}} \in \mathbb{R}^{Q \times k_{\M}}$, as the \emph{functional} or \emph{spectral} regressor. A visualization of the functional regressor is reported in Figure \ref{fig:regressor}; each column represents a dimension of the LBO basis (one of the eigenfunctions), and each row a different joint. Further comments and analysis are shown in Section \ref{sec:insight}.

\vspace{0.5cm}\noindent\textbf{Relation between spectral and standard regressor.}
Let us consider a single shape $\mathcal{M}$ with its 3D coordinates $X_{\M}$, its LBO eigenfunctions $\bm{\Phi}_{k_{\M}}$, and the Fourier coefficients of its coordinates $\widehat{X}_{\M}$.
From Equation \ref{eq:specregrX} and from $\bm{J}_{\mathcal{M}} = \bm{R}X_{\M}$, we can write:
\begin{equation}
\label{eq:relation}
    \bm{J}_{\mathcal{M}} = \bm{R}X_{\M} \approx \widehat{\bm{R}} \widehat{X}_{\M} = \widehat{\bm{R}} \bm{\Phi}^{\dagger}_{k_{\M}} X_{\M},
\end{equation}
from which we obtain:
\begin{equation}
    \bm{R} \approx \widehat{\mathbf{R}} \bm{\Phi}^{\dagger}_{k_{\M}}.
\end{equation}
From the definition of the Moore–Penrose pseudo-inverse we have:
\begin{equation}
    \bm{R}\bm{\Phi}_{k_{\M}} \approx \widehat{\bm{R}}\underbrace{ \bm{\Phi}_{k_{\M}}^{\dagger}\bm{\Phi}_{k_{\M}}}_{Id_{k_{\M} \times k_{\M}}} = \widehat{\bm{R}},
\end{equation}
and finally we can derive these two relations:
\begin{align}
\label{eq:approx1}
\bm{R} &\approx \widehat{\bm{R}} \bm{\Phi}^{\dagger}_{k_{\M}} \\
\label{eq:approx2}
\widehat{\bm{R}} &\approx \bm{R}\bm{\Phi}_{k_{\M}}.
\end{align}

\subsection{Skeleton transfer algorithm}

\begin{algorithm}
    \caption{Skeleton Transfer}\label{euclid}
    \begin{algorithmic}[1]
    \State Input: A rigged shape $\mathcal{M}$, a target shape $\mathcal{N}$, and a set of seven landmarks
    \State Step 1: Estimate the spatial regressor $\bm{R}$ by optimizing the energy $E$
    \State Step 2: Compute the FMAP $\bm{C}$ using the given landmarks
    \State Step 3: Convert the spatial regressor $\bm{R}$ in the functional representation $\widehat{\bm{R}}$
    \State Step 4: Compute $\bm{J}_{\mathcal{N}}$ using the transferred spectral regressor 
    \State Output: A skeleton $\bm{J}_{\mathcal{N}}$ for the target shape $\mathcal{N}$
    \end{algorithmic}
    \label{alg:pipeline}
\end{algorithm}

We consider a model $\mathcal{M}$ that is animation-ready, composed by a triangular mesh represented by $n_{\mathcal{M}}$ vertices $X_{\M}\in \mathbb{R}^{n_{\mathcal{M}}\times 3}$ and equipped with a skeleton $\bm{J}_{\mathcal{M}}\in \mathbb{R}^{Q \times 3}$ composed by $Q$ joints.
The SMPL model proposed in \cite{loper2015smpl} is one possible model that we consider in our experiments.
Our objective is to transfer the rigging of $\mathcal{M}$ to a new shape $\N$ represented by a triangular mesh with $n_{\mathcal{N}}$ vertices $X_{\N} \in \mathbb{R}^{n_{\mathcal{N}}\times 3}$. As additional information, we require some landmarks that for the human case are seven. Five of them are located on protrusions and can be automatically computed exploiting the method proposed in \cite{marin2020farm}. The remaining two are on the front and the back of the torso. In our experiments, we select them manually since this requires just a few seconds.

\vspace{0.5cm}\noindent\textbf{Step 1: Spatial Regressor Optimization}
We want to compute the mesh skeleton through a linear operator $\widehat{\bm{R}}$ defined over the Fourier coefficients of the 3D coordinates of the vertices of the input mesh. To do so, we first obtain a standard regressor $\bm{R}$ as the solution of an optimization defined as the linear combination of 4 energies. 

\paragraph{Reconstruction Term.} 
The first energy asks that the regressor $\bm{R}$ returns the correct position of the joints:
\begin{equation}
    E_{rec} = \left\Vert \bm{R}X_{\M} - \bm{J}_{\M} \right\Vert^2 .
\end{equation}
\paragraph{Locality Term.}
To have a regressor invariant to pose, each joint should depend only on the vertices linked by a rigid relation to it, i.e. by a near set of them. To do so, we define a mask matrix $F \in \mathbb{R}^{Q\times n_{\mathcal{M}}}$, that has $F(q,i) = 1$ if the position of the vertex $v_i$ influences the position of the joint $J_{q}$ and $0$ otherwise. To select which vertices should affect the joint position, we took the ones in the intersection between the non-zero skinning weights belonging to each joint and its hierarchical parent. The resulting set is near to the joint, and so we selected them to populate $F$. We decided to choose the vertices between the joint and its parent since they are near the joint and they have low weights so the deformation of these vertices is low changing between a pose and another. 
Then, the energy is:
\begin{equation}
    E_{loc} = \left\Vert \bm{R}\odot F - \bm{R} \right\Vert^2 
\end{equation}
where $\odot$ is the element-wise product.
\paragraph{Sparsity Term.}
While $E_{loc}$ guarantees that joints are affected by a limited set of vertices, nothing has been imposed on the entries of $R$ that correspond to zero values in the mask $F$. For this reason, we promote zero-values in all these entries that should not contribute to the computation of $\bm{J}_{\M}$:
\begin{equation}
E_{spa} = \|  |F - \mathbf{1}| \odot \bm{R} \|^{2},
\end{equation}
where $\mathbf{1}$ is a matrix with the same size of $F$, the entries of which are all equal to $\mathbf{1}$.
\paragraph{Convexity Term.}
We observed in our experiment that requiring the weights of each joint to sum to 1 helps to obtain joints inside the mesh boundaries, even if we do not explicitly require that all the weights are non negative. We implement this through the following energy:
\begin{equation}
    E_{con}  = \left\Vert \bm{R}e_{n_{\M}} - e_Q \right\Vert^2
\end{equation}
where $e_{n_{\M}}$ and ${e_Q}$ are vectors of ones with length $n_{\M}$ and the number of joints respectively. 
\paragraph{Optimization formulation.}
The optimization to recover the desired regressor is obtained by a linear combination of these four energies:
\begin{equation}
    E = \omega_{rec} * E_{rec} + \omega_{loc} * E_{loc}  + \omega_{spa} * E_{spa} +  \omega_{con} * E_{con}
\end{equation}
where $\omega_{rec}$, $\omega_{loc}$, $\omega_{spa}$, $\omega_{con}$ are positive real numbers that act as weights. We empirically set them to $\omega_{rec} = 1$, $\omega_{loc}= 10e3$, $\omega_{con}=1$ and $\omega_{spa}=100$. Putting high weights on $E_{loc}$ and $E_{spa}$ forces the regressor to better compute the joints location as a function of the vertices selected by $F$. These dependencies between each joint and the vertices are independent across the different joints. Apart from setting a rough balance between energies, no exhaustive research has been done.
\paragraph{Implementation.}
We use the Tensorflow auto-differentiation library to solve this problem, relying on the Adam optimizer with a learning rate of $0.1$ and using 1000 iterations.

\vspace{0.5cm}\noindent\textbf{Step 2: Functional Maps computation}
Given a new shape $\mathcal{N}$ for which the rigging is not available, our goal is to transfer the regressor from $\M$ to $\N$.  Hence, we estimate a functional map $\bm{C}$ between $\mathcal{F}(\M)$ and $\mathcal{F}(\N)$ following the procedure proposed in \cite{nogneng2017informative}. We set both the sizes of the Fourier bases $k_{\M}$ and $k_{\N}$ computing an initial functional map of dimension $20 \times 20$. We refine this map with ZoomOut the refinement technique proposed in \cite{zoomout}, bringing to a $120 \times 120$ map. Thanks to our functional definition of the regressor, we are able to perform this transfer without requiring a pointwise correspondence between $\M$ and $\N$.

\vspace{0.5cm}\noindent\textbf{Step 3: Regressor conversion}
\label{sec:skeltransf}
Once we optimize the regressor $\bm{R}$ we can easily compute its functional representation $\widehat{\bm{R}}$ exploiting Equation \ref{eq:approx2}:
$\widehat{\bm{R}} \approx \bm{R}\bm{\Phi}_{k_{\M}}$.

\vspace{0.5cm}\noindent\textbf{Step 4: Skeleton transfer}
Finally, the skeleton of the shape $\mathcal{N}$ is obtained by:
\begin{equation}
     \bm{J}_{\mathcal{N}} = \widehat{\bm{R}} \bm{C} \widehat{X}_{\N}.
\end{equation}

Our complete implementation is available online for full reproducibility.
\footnote{Demo Code: \href{GitHub Repository}{\url{https://github.com/PietroMsn/Functional-skeleton-transfer}} }
\section{Results}
\label{sec:results}

In this section, we test the robustness of our method on several shapes collected from different datasets, providing both qualitative and quantitative evaluations and comparisons to the most related approaches. Moreover, we analyze the computational efficiency of our approach. 

\begin{wrapfigure}[17]{R}{0.33\linewidth}
\vspace{-1.0cm}

  \begin{overpic}
  [trim=0cm 0cm 0cm 0cm,clip,width=\linewidth]{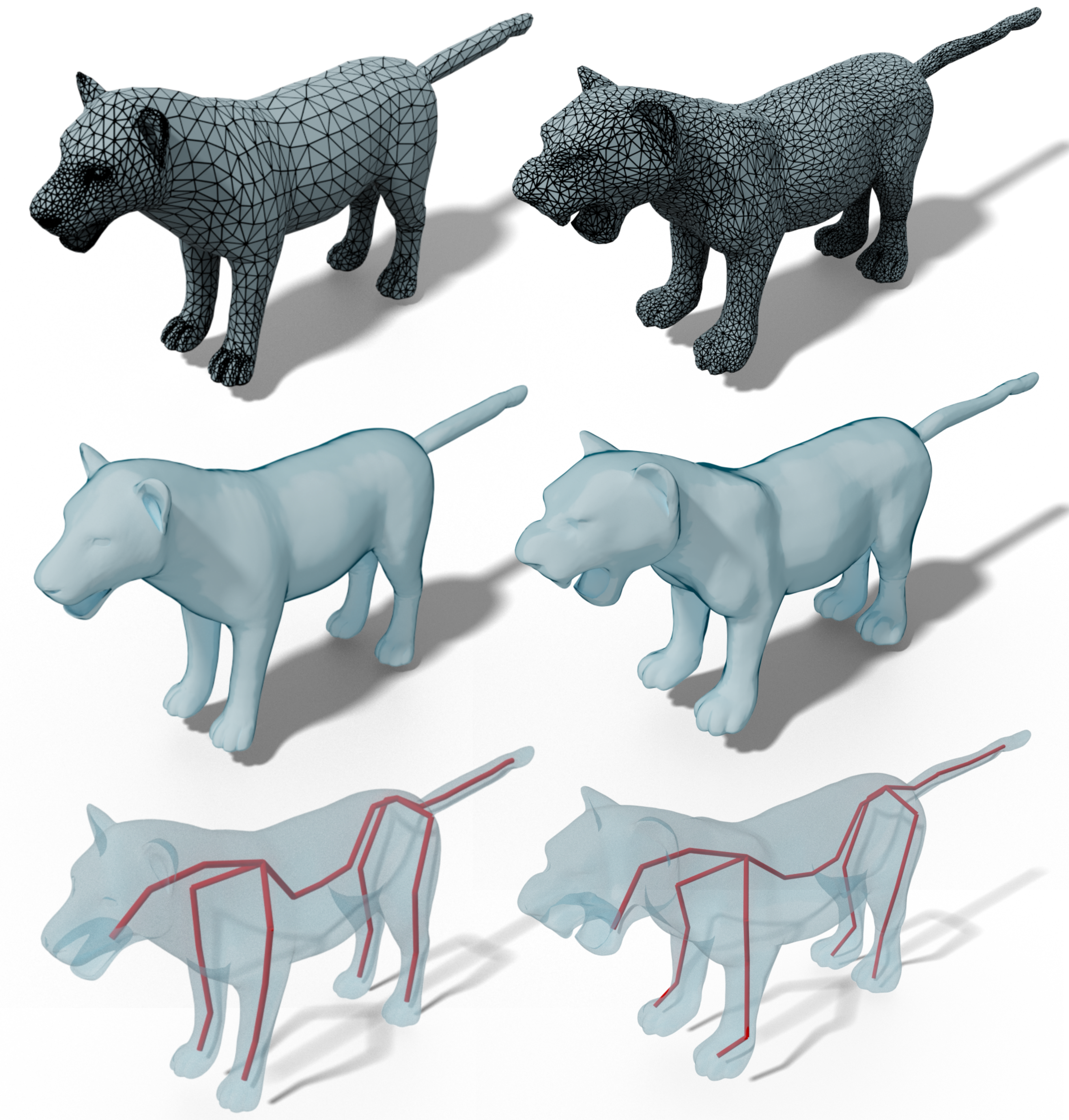}
  \put(60,55){}
  \end{overpic}
  \caption{\label{fig:smal} A transfer between two animals with different geometry and discretization. Notice that finding a point-to-point correspondence, in this case, could be difficult even if the two shapes are in the same pose.}
\end{wrapfigure}

\vspace{0.5cm}\noindent\textbf{Data.} We evaluate our method on a wide selection of shapes. From the \emph{SHREC19} dataset \cite{SHREC19}, which consists of 44 shapes of different subjects in many poses and with varying discretizations, and the \emph{Dyna} set, composed of 155 shapes of 8 different subjects from Dynamic Faust \cite{dfaust}. They present the typical artifacts of real scans and are significantly different in their body characteristics. We also stress our transferring by assuming that Dyna is composed just of point clouds (i.e., without a triangulation) and by estimating the appropriate LBO operator from \cite{Sharp:2020:LNT}. Notice that this is a significantly challenging setting (marked with a final \emph{P.C.} in the table), showing our flexibility. Finally, we also consider two different populations of characters to show the generalization of our method: animals (i.e., \emph{SMAL}\cite{smal}; examples are reported in Figures \ref{fig:teaser} and \ref{fig:smal}) and children (i.e., \emph{SMIL}\cite{Hesse2018}; examples are reported in Figures \ref{fig:teaser}). 
We generate ten different shapes from each of the respective morphable models introducing an independent remeshing procedure to break the regularity of the connectivity.

\begin{figure}[t]  
\setlength{\tabcolsep}{2pt}
\begin{tabular}{l r r}
  \centering
\begin{minipage}{0.4\linewidth}
  \begin{tabular}{|l|c|c|c|}
\hline
\footnotesize{Dataset} & \footnotesize{mean} & \footnotesize{min} & \footnotesize{max} \\ 
\hline 
\footnotesize{SHREC19 functional} & \footnotesize{0.0210}
 & \footnotesize{0.0072} & \footnotesize{0.0512} \\ 
\footnotesize{SHREC19 pointwise}  & \footnotesize{0.0215}
 & \footnotesize{0.0072} & \footnotesize{0.0525} \\
\hline 
\footnotesize{Dyna functional P.C.} & \footnotesize{0.0237} & \footnotesize{0.0078} & \footnotesize{0.1934}\\
\footnotesize{Dyna pointwise P.C.}  & \footnotesize{0.0255} & \footnotesize{0.0125} & \footnotesize{0.2013} \\
\hline 
\footnotesize{SMIL functional} & \footnotesize{0.0526}
 & \footnotesize{0.0376} & \footnotesize{0.0658} \\ 
\footnotesize{SMIL pointwise}  & \footnotesize{0.0527}
 & \footnotesize{0.0376} & \footnotesize{0.0660} \\
\hline 
\footnotesize{SMAL functional} & \footnotesize{0.0635} & \footnotesize{0.0323} & \footnotesize{0.1190} \\
\footnotesize{SMAL pointwise} & \footnotesize{0.0692} & \footnotesize{0.0350} & \footnotesize{0.1323} \\
\hline 
\end{tabular}
\end{minipage}
&
\hspace{0.5cm}
\begin{minipage}{0.25\linewidth}
  \begin{overpic}
  [trim=0cm 0cm 0cm 0cm,clip,width=1\linewidth]{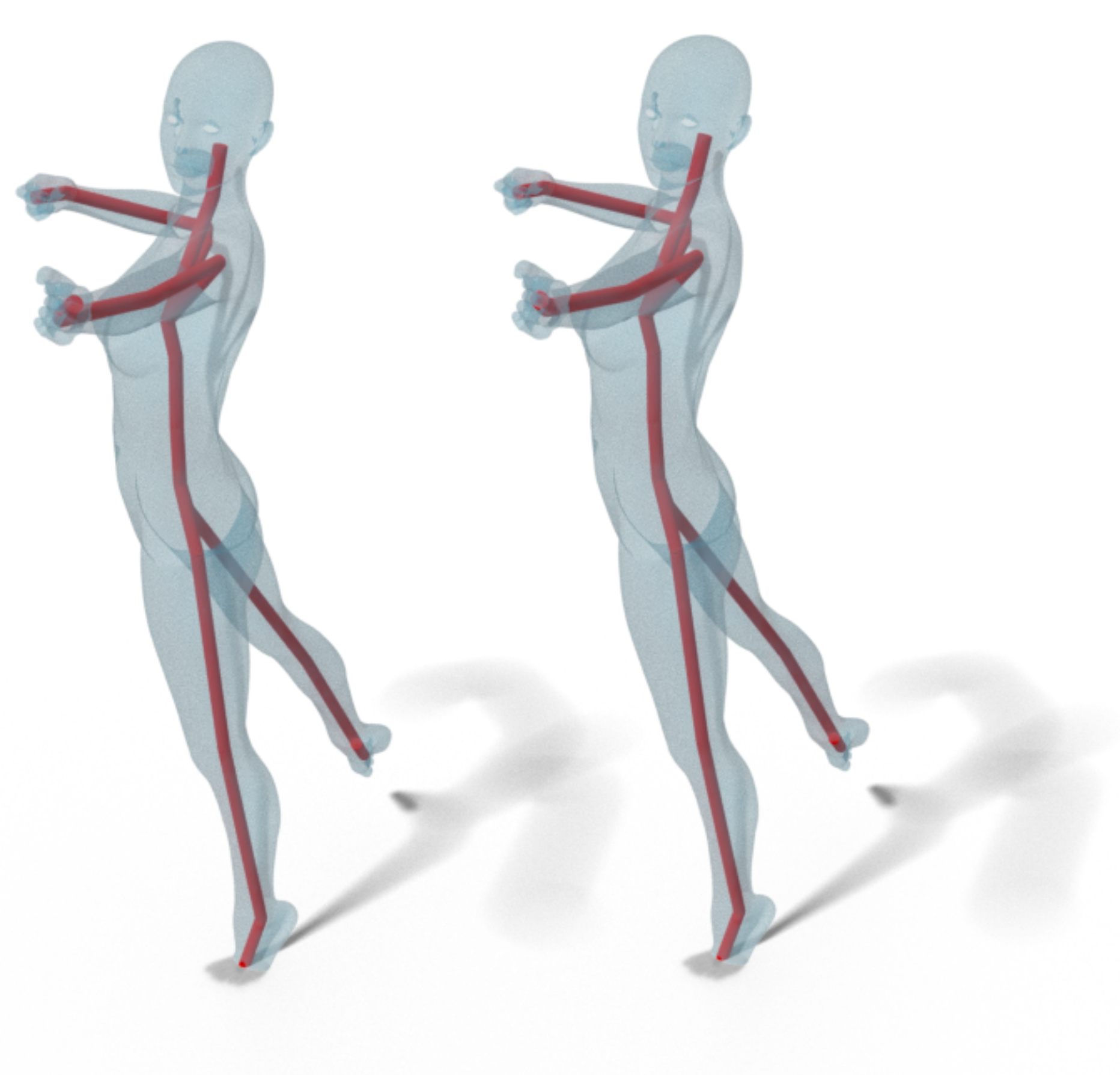}
  \put(8,-3){\footnotesize{functional}}
  \put(55,-3){\footnotesize{pointwise}}
  \end{overpic}
\end{minipage}
&
\hspace{-0.5cm}

\begin{minipage}{0.25\linewidth}
  \begin{overpic}
  [trim=0cm 0cm 0cm 0cm,clip,width=1\linewidth]{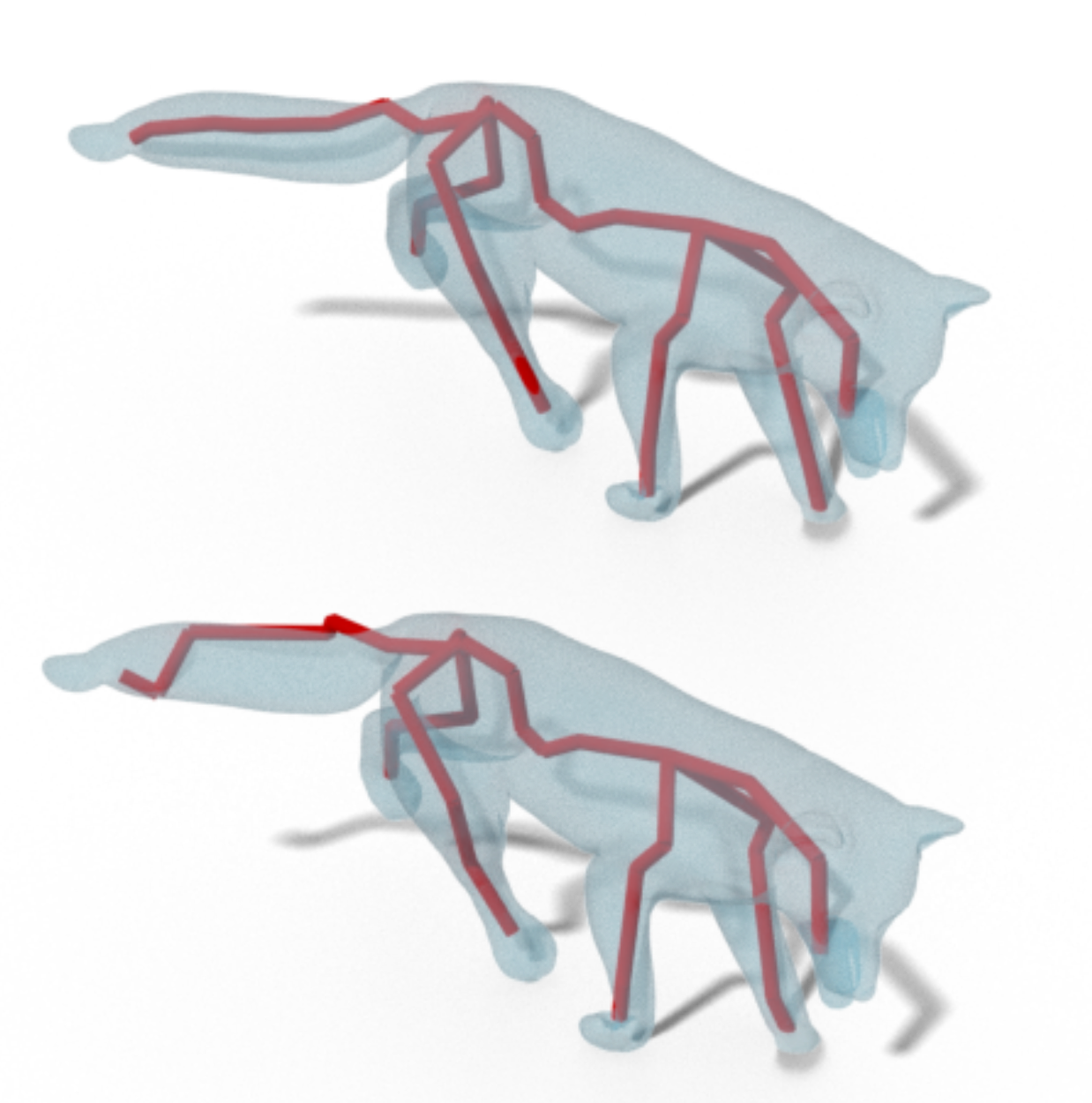}
  \put(30,-1){\footnotesize{pointwise}}
  \put(30,48){\footnotesize{functional}}
  \end{overpic}
\end{minipage}
\end{tabular}
\vspace{0.2cm}

\captionof{table}{\label{tab:errors}Mean, minimum and maximum error averaged over the datasets, comparing functional and spatial representations. All the errors are expressed in meters. On the right, two examples of functional and pointwise transfer. In some cases, the difference is minimal, while the functional approach requires a step less; on the animal the functional transfer is noticeably better than the one induced by the pointwise correspondence.}
\vspace{-0.5cm}
\end{figure}

\vspace{0.5cm}\noindent\textbf{Transferring.}  We quantitatively evaluated our method on the skeleton transfer task, and we report the results in Table \ref{tab:errors}. For each dataset, the transfer is performed i) by using the functional approach described in Section \ref{sec:skeltransf} (\emph{functional} in the table), and ii) by converting the functional mapping to a point-to-point correspondence and using regressor in its spatial form (\emph{pointwise} in the table). In the first two datasets of adult humans (SHREC19 and Dyna) the shapes shared the same skeleton of the SMPL model \cite{loper2015smpl}, while for children and animals the skeleton is provided by the related morphable models (SMIL and SMAL). For each setting, we report the Mean Squared Error with respect to the skeletons obtained using ground truth correspondences with the related morphable model. 

%

\begin{figure}
  \begin{overpic}
  [trim=0cm 0cm 0cm 0cm,clip, width=0.8\linewidth]{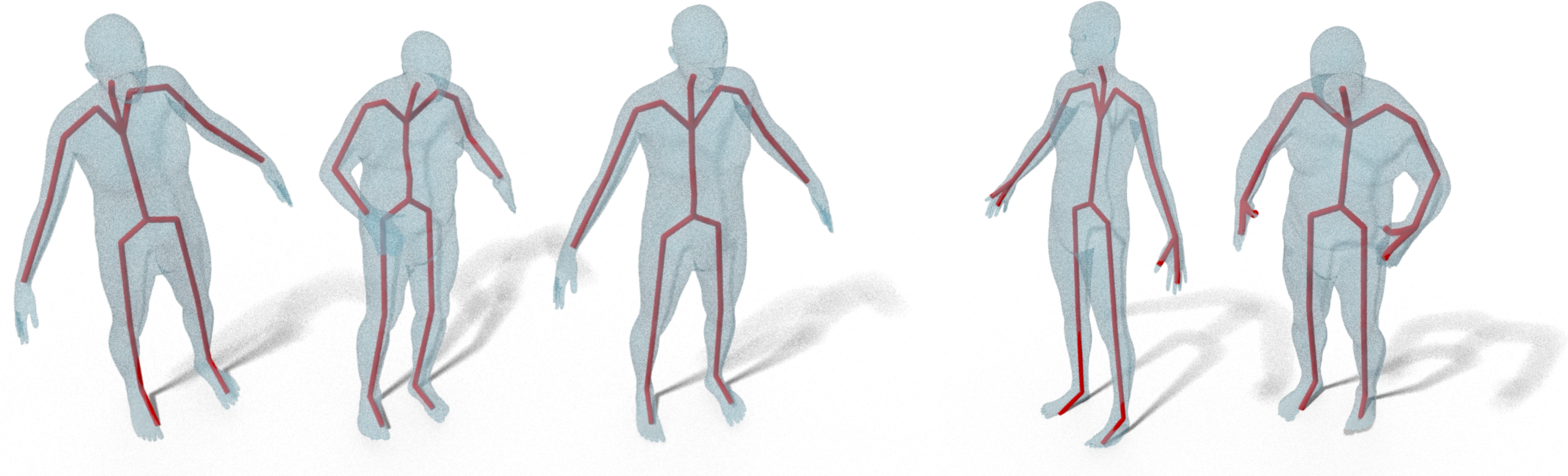}
  \put(20,-1){$22$ Joints}
  \put(70,-1){$34$ Joints}
  \put(6,32){\tiny{Src}}
  \put(23,32){\tiny{Src $\rightarrow$ Tar}}
  \put(38,32){\tiny{Src $\rightarrow$ Tar $\rightarrow$ Src}}
  \put(68,32){\tiny{Src}}
  \put(80,32){\tiny{Src $\rightarrow$ Tar}}
  \end{overpic}
  %
  \caption{\label{fig:mixamo} Transfer examples using two different skeletons from Mixamo. On the left, a version with $22$ joints; on the right, one composed of $34$ joints. Please note the additional joints in the hands region on the second pair. We remark the coherence and independence from the involved skeleton of the two results.}
\end{figure}

\begin{figure}
  \begin{overpic}
  [trim=0cm 0cm 0cm 0cm,clip, width=0.8\linewidth]{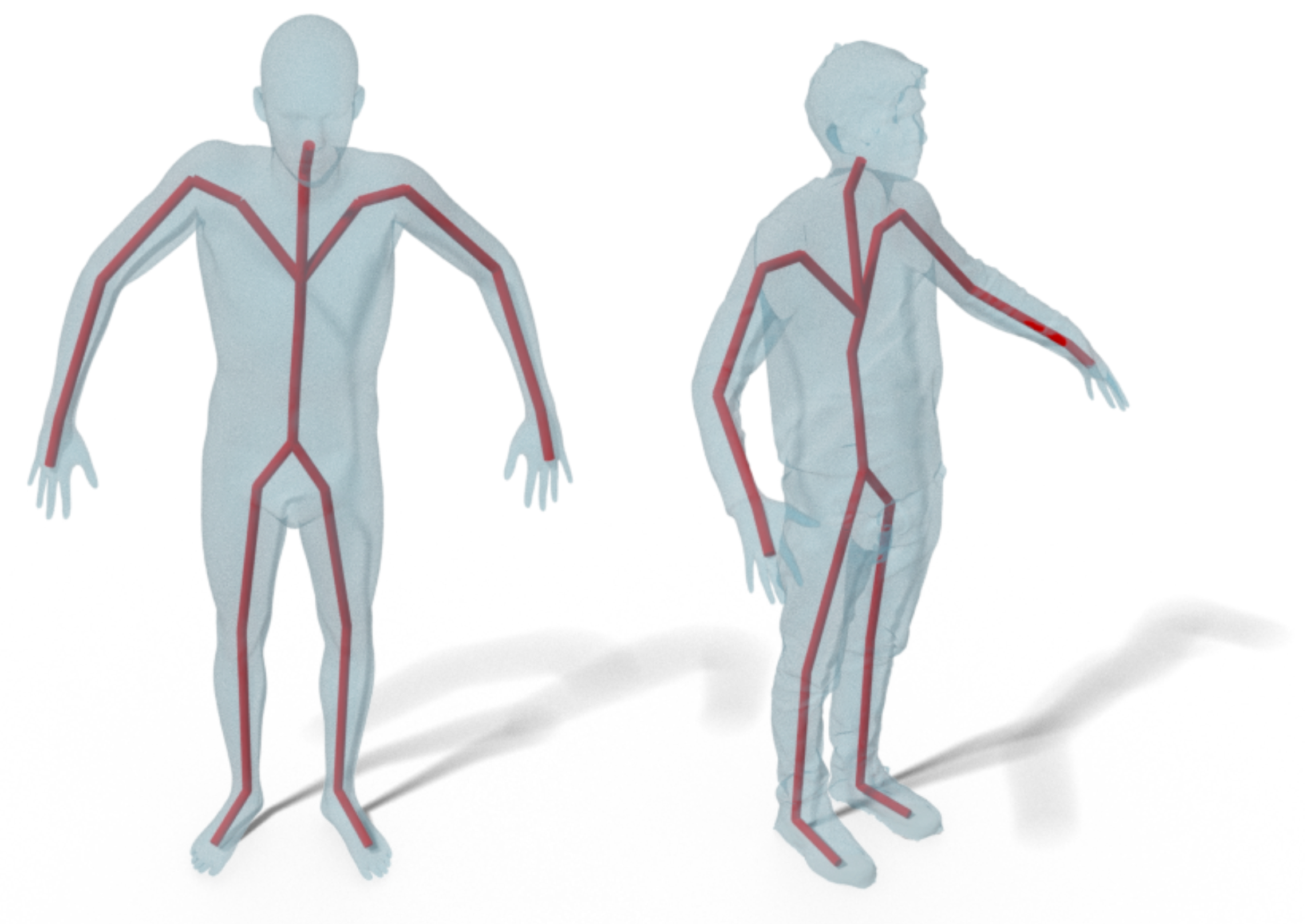}
  \put(44,1){}
  \put(72,1){}

  \end{overpic}
  %
  \caption{\label{fig:clothes}A transfer of a skeleton between a FAUST shape and a clothed model. Although the differences in the surface of the two shapes, our regressor can transfer the skeleton on the clothed model.}
\end{figure}

We would remark that the dense matching is a step built upon the functional approach (see Section \ref{sec:fmaps}); it cannot be obtained without computing the functional maps before.
Our method obtains stable results, despite the variabilities. The results achieved by our functional approach are similar to the ones obtained by applying the SMPL regressor to the point-to-point map obtained by functional map conversion, showing that the latter is not required. Solving for a point-to-point correspondence is challenging for these shapes, and there is not always an exact solution. We report a result on SMAL in Figure \ref{fig:smal}, highlighting the difference in their connectivity. Also, in Figure \ref{fig:mixamo}, we tested our method on two different skeletons from Mixamo \footnote{\url{www.mixamo.com}}, one with $22$ joints and the extended version with $34$ joints that also control the hands. We show that our method can be applied to different skeletons and that we can represent the joints on terminal protrusions (i.e., hands) without impacting the global results on the rest of the shape. Furthermore, in Figure \ref{fig:clothes} we tested our resilience to the presence of surface noise, by transferring the skeleton to a shape with clothes and hair. The qualitative result is convincing.

\begin{wrapfigure}[11]{R}{0.32\linewidth}
\setlength{\tabcolsep}{1pt}
\vspace{-1.5cm}

\begin{tabular}{|l|c|c|c|}
\hline
\footnotesize{Method} & \footnotesize{mean} & \footnotesize{min} & \footnotesize{max} \\ 
\hline 
\footnotesize{Dyna Our}  & \footnotesize{0.0278} & \footnotesize{0.0076} & \footnotesize{0.0817} \\
\footnotesize{Dyna FARM}  & \footnotesize{0.0332} & \footnotesize{0.0096} & \footnotesize{0.0815} \\ 
\footnotesize{SHREC19 Our}  & \footnotesize{0.0210} & \footnotesize{0.0072} & \footnotesize{0.0512} \\ 
\footnotesize{SHREC19 FARM}  & \footnotesize{0.0295} & \footnotesize{0.0005} & \footnotesize{0.1277} \\ 
 \hline 
\end{tabular}

\captionof{table}{Comparison with \cite{marin2020farm} for the skeleton transfer from SMPL shape to Dyna and SHREC19 shapes. The average, minimum and maximum errors are expressed in meters and computed as the distance between the estimated and the SMPL regressor joints.}\label{tab:farm_comp}
\end{wrapfigure}

\begin{figure}
  \begin{overpic}
  [trim=0cm 0cm 0cm 0cm,clip, width=0.8\linewidth]{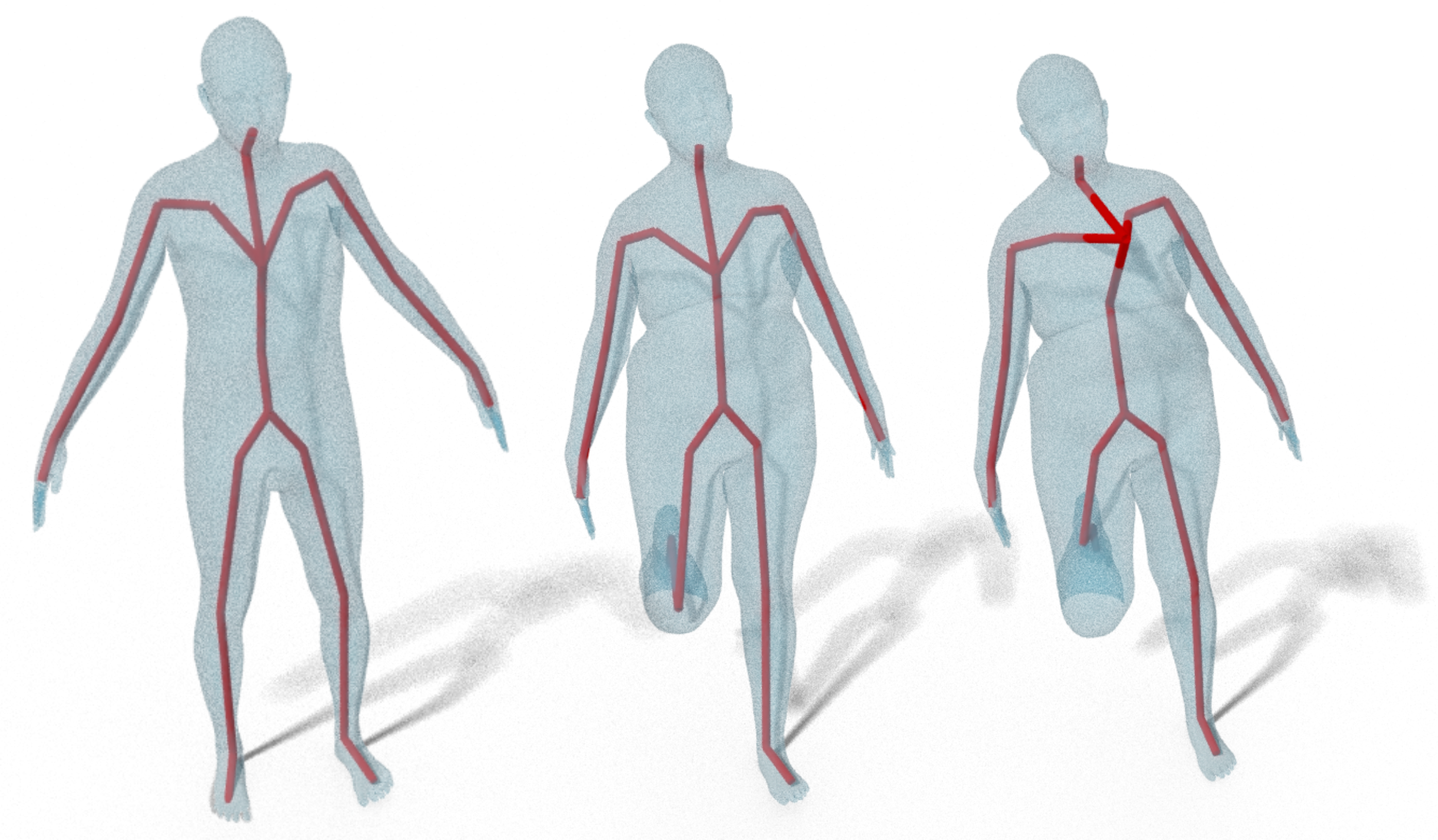}
  \put(48,0){Our}
  \put(75,0){\cite{avril2016animation}}

  \end{overpic}
  %
  \caption{\label{fig:comp}Comparison between our method and the regressor presented in \cite{avril2016animation}.}
\end{figure}

\vspace{0.5cm}\textbf{Comparison.}  We compare our skeleton transfer approach against the method proposed in \cite{marin2020farm}, which we consider our baseline. This method consists of transferring the coordinates of the target into the source space and then applying the spatial regressor to the low-pass representation. While the two approaches follow similar principles, \cite{marin2020farm} applies the spatial regressor to a degraded representation of the target mesh. This combination of a high-frequency operator with low-frequency structure generates undesirable artifacts, and it cannot be considered useful for animation. In Table \ref{tab:farm_comp} we report a comparison on the adult human datasets, showing an improvement of more than $20\%$. Finally, in Figure \ref{fig:comp}, we compare ourselves against \cite{avril2016animation}. To the best of our knowledge, the code of \cite{avril2016animation} is not publicly available. We thus implement their regressor and apply it to the test shape using the point-to-point mapping obtained by the functional map involved in our method. Our method outputs better results, in particular in the chest region of the shape. The reason is that \cite{avril2016animation} assume the two shapes are in the same pose, while our do not.

\subsection{Efficiency and timing}
\begin{figure}[t]  
\begin{center}
\setlength{\tabcolsep}{1pt}
\begin{tabular}{c r}
  \centering
  \hspace{1cm}

\begin{minipage}{0.5\linewidth}

\begin{tabular}{|c|c|c|c|}
\hline
\footnotesize{num. vert.} & \footnotesize{mapping} & \footnotesize{regressor} & \footnotesize{total} \\ 
\hline 
\footnotesize{1K}  & \footnotesize{10.33 s} & \footnotesize{11.43 s} & \footnotesize{24.06 s}\\
\footnotesize{2K}  & \footnotesize{12.63 s} & \footnotesize{12.65 s} & \footnotesize{27.53 s}\\
\footnotesize{3K}  & \footnotesize{13.05 s} & \footnotesize{12.43 s} & \footnotesize{26.68 s}\\ 
\footnotesize{5K}  & \footnotesize{12.91 s} & \footnotesize{11.44 s} & \footnotesize{26.67 s}\\ 
\footnotesize{10K}  & \footnotesize{17.23 s} & \footnotesize{12.09 s} & \footnotesize{31.64 s}\\ 
\footnotesize{30K}  & \footnotesize{40.82 s} & \footnotesize{14.86 s} & \footnotesize{57.90 s}\\ 
 \hline 
\end{tabular}

\end{minipage}
&
  \hspace{-1cm}
\begin{minipage}{0.5\linewidth}
  \begin{overpic}
  [trim=0cm 0cm 0cm 0cm,clip,width=1\linewidth]{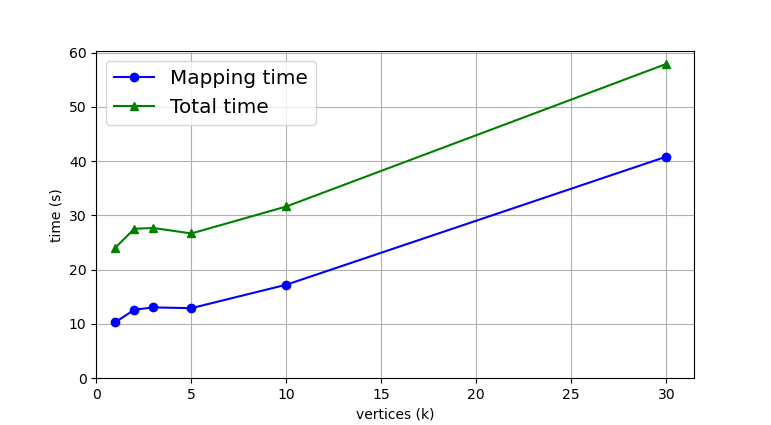}
  \put(60,55){}
  \end{overpic}
\end{minipage}
\end{tabular}
\end{center}
\caption{\label{fig:timing} On the left, we report the execution time of the two main steps of our method with respect to an increasing number of vertices in the target shape. On the right, we highlight the linearity in this relation.}
\vspace{-0.5cm}
\end{figure}

Since efficiency is a critical aspect of skeleton transfer, we measure our method timing on a subset of 54 shapes from Dynamic Faust subsampled with seven different vertex resolutions: 1K, 2K, 3K, 5K, 10K, and 30K. 
As can be noticed in Figure \ref{fig:timing}, our method execution time depends on the mapping and is linear on the number of the vertices of the shapes, while other operations have almost a constant cost. In all cases, the overall execution time is under 1 minute, even for 30K resolution meshes. The machine used for this experiment is composed of a Ryzen 7 with a frequency of 3.6 GHz and 64 Gb of RAM. We consider our setting comparable to one of previous methods \cite{avril2016animation} \cite{moutafidou2019mesh} which relies on a point-to-point correspondence computation; even if we cannot perform a direct comparison due to lack of available code, our method runs significantly faster on shapes of similar resolution (from 2$\times$ to 10$\times$). We remark that we also require less user intervention.
%
%
\subsection{Skinning weights transfer}
While our work focuses on skeleton transfer, we also wondered about our capability to transfer skinning-weights information.

\vspace{0.5cm}\noindent\textbf{LBS principles} 
Given a shape $\M$ with $n_{\M}$ vertices, the Linear Blend Skinning (LBS) on $\M$ is defined by a set of components that give rise to a semantic and kinematic meaningful deformation of the 3D object represented by the shape $\M$ (e.g. the possible and realistic deformations of a human body). LBS is one of the most popular frameworks for skinning characters due to its efficiency and simplicity. 
The key idea is defining a relation between each vertex and all the joints of a skeleton as a scalar weight. Hence, the position of a vertex is a weighted sum of the contribution of all the rotations of all the joints:
\begin{equation}
\bar{v_i}  = \sum_{q=1}^Q w_{i,q} R^\prime (\mathbf{\theta}, \bm{J}) v_i,
\end{equation}
Where the transformation $R^\prime (\mathbf{\theta}, \bm{J})$ takes into account the transformation induced by the hierarchy in the kinematic tree:

\begin{equation}
    R_q^{\prime} (\mathbf{\theta}, \bm{J})  =  R_q(\mathbf{\theta}, \bm{J}) R_q(\mathbf{\theta}^{*},\bm{J})^{-1}
\end{equation}
\begin{equation}
        R_j(\mathbf{\theta}, \bm{J})  =  \prod\limits_{l \in A(J_q)} \left[
\begin{array}{c|c}
\text{rodr}(\theta_l) & j_l \\
\hline
0 & 1
\end{array}
\right]
\end{equation}

\noindent Where $\text{rodr}$ represent the Rodrigues formula to convert axis-angle notation to rotation matrix; $j_l$ is the single joint center; $A(J_q)$ is the list of ancestors of $J_q$ in the kinematic tree; $R_q(\mathbf{\theta}^{*},\bm{J})$ is the transformation of joint $q$ in the world frame, and $R_q^\prime (\mathbf{\theta}, \bm{J})$ is the same after removing the transformation induced by the rest pose $\mathbf{\theta}^{*}$. We refer to \cite{loper2015smpl} for further details.

\vspace{0.5cm}\noindent\textbf{Skinning transfer}
 Transferring the skinning weights consists of moving some high-frequencies details not well represented by the first LBO eigenfunctions. In particular, by using the functional map $C$, we observe the typical Gibbs phenomenon of Fourier analysis over all the surface which makes the skinning weights on the target too noisy for being usable. However, while a full investigation is beyond our scope, we would suggest a first possible solution. First, we transfer the coordinate functions of the target using our $C$, mapping them in the SMPL space. This generates a low-pass representation of the target geometry with the connectivity of SMPL. Then, we compute the Euclidean nearest-neighbour in the $3$D to transfer the skinning weights defined on SMPL to the target low-pass representation. In this way, the Gibbs phenomenon affects the coordinate, while the nearest-neighbour assignment preserves the skinning weights original values. The results can be appreciated in the \emph{attached video}, where we show an animation transferred from a SMPL shape to a new target shape (a remeshed shape from the FAUST dataset \cite{faust}) using the transferred skeleton and skinning. We consider this an exciting direction, which elicits further analysis. 

\begin{figure}
 \begin{center}
  \begin{overpic}
  [trim=0cm 0cm 0cm 0cm,clip, width=0.8\linewidth]{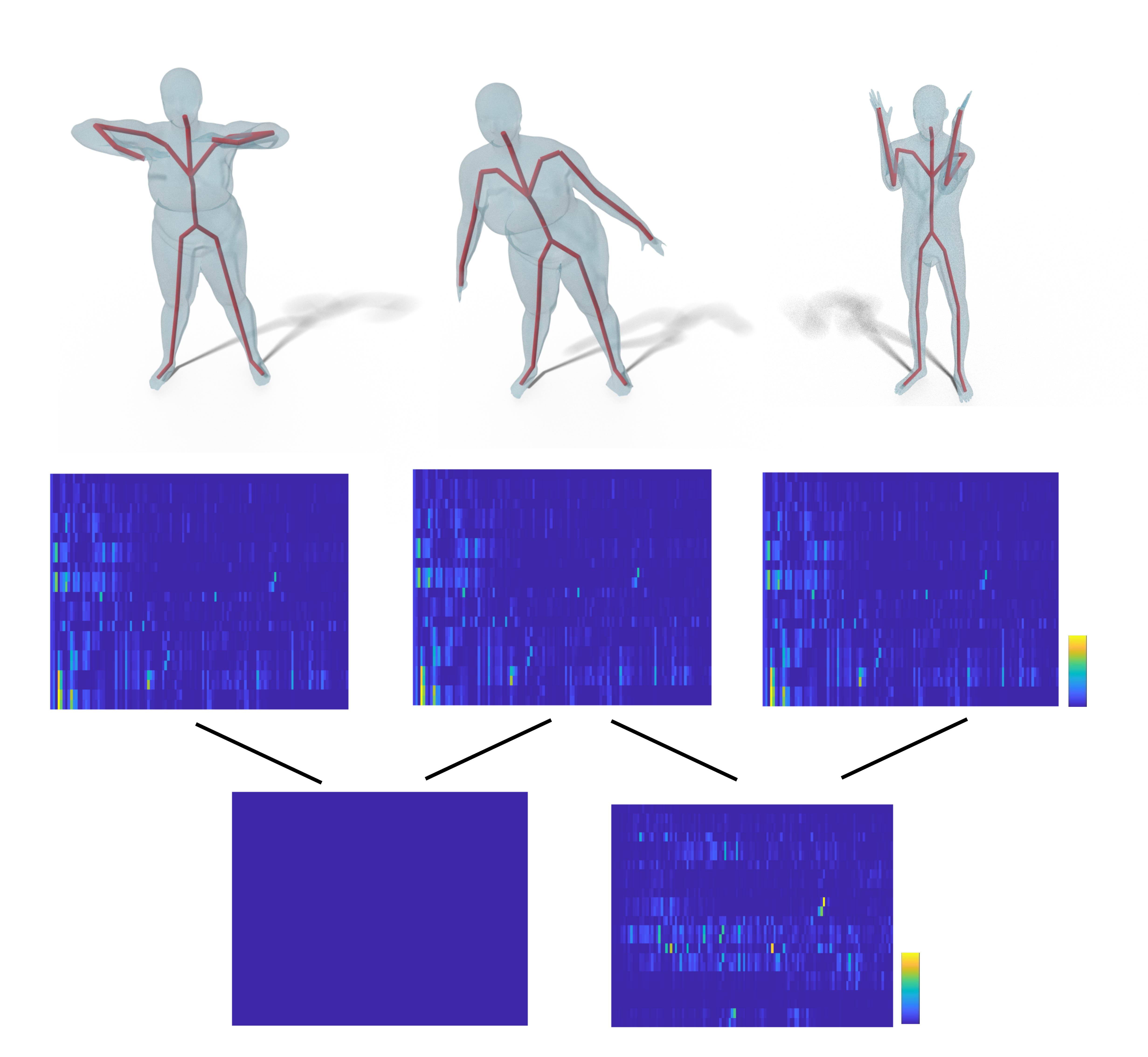}
  \put(15,87){\tiny{$A$}}
  \put(47,87){\tiny{$B$}}
  \put(78.5,87){\tiny{$C$}}
  \put(15,53.5){\tiny{$\widehat{\bm{R}}_A^2$}}
  \put(47,53.5){\tiny{$\widehat{\bm{R}}_B^2$}}
  \put(78.5,53.5){\tiny{$\widehat{\bm{R}}_C^2$}}
  \put(28,-2){\tiny{$|\widehat{\bm{R}}_B - \widehat{\bm{R}}_A|$}}
  \put(59.5,-2){\tiny{$|\widehat{\bm{R}}_B - \widehat{\bm{R}}_C|$}}
  \put(95.3,35){\tiny{6}}
  \put(95.3,29){\tiny{0}}
  \put(81,7.5){\tiny{0.2}}
  \put(81,1.5){\tiny{0}}
  \end{overpic}
\end{center}
\caption{\label{fig:diff} A visualization of the skeleton transfer on three shapes: $A, B, C$. In the first row: the obtained results. In the second row: the spectral version of the regressor transferred on the meshes; we squared their values to ignore sign flips of the eigenfunctions. In the third row, we compute the differences between two shapes which share the same identity, and two shape which not. Notice that in the second case it can be noticed a marked difference.}
\end{figure}

\section{Spectral regressor insights}
\label{sec:insight}
We would emphasize some of the properties of our spectral regressor, which could open it to impactful and innovative applications. 
To highlight the intrinsic properties of our representation, we show in Figure \ref{fig:diff} some results of our regressor transfer. The three shapes A, B and C in the first row are significantly non-isometric to the source shape (depicted on the top left of Figure \ref{fig:pipeline}).
In the second row, we report the pointwise square values of our functional regressor transferred to these three shapes from the source shape. More explicitly, given $\bm{C}_A$, $\bm{C}_B$ and $\bm{C}_C$, the functional maps between the source and the three shapes are: $\widehat{\bm{R}}_A = \widehat{\bm{R}} \bm{C}_A$, $\widehat{\bm{R}}_B = \widehat{\bm{R}} \bm{C}_B$ and $\widehat{\bm{R}}_C = \widehat{\bm{R}} \bm{C}_C$ respectively. In these matrices, the rows correspond to the joints and the columns to the Fourier basis functions. 
At first sight, some patterns emerge. The joints from the central region of the body (corresponding to the top rows of the matrices) mainly influence the low frequencies (i.e. the first 20-30 columns of the matrix).  
Instead, the joints from the limbs (the bottom rows of the matrix) involve also high frequencies. 
The matrices look similar, highlighting a shared structure. However, there are differences between transfers on the same subject (left and middle shapes) and transfer between different identities (the right shape).
We visualize these differences in the bottom part of the image. We see how the first two are almost identical, while non-isometric shapes present differences, especially in high frequencies

We think our intrinsic operator can extract a compact representation of the extrinsic embedding of the shape (i.e. the pose). We believe that it could be beneficial for several pipelines, filling the gap generated by the missing extrinsic information in the intrinsic analysis, as occurs, for example, for the functional maps and the shape difference operators where the skeleton (extrinsic) information could help to solve the standard issues related to the (intrinsic) symmetries of shapes \cite{MapTree}. 
We propose the exploration of shape collections and the learning of a latent representation in a data-driven pipeline as two interesting future directions.

\section{Conclusions}
\label{sec:conclusions}

\vspace{0.5cm} In this paper, we have shown a straightforward method to transfer skeleton information across shapes efficiently. Our functional approach is agnostic to the mesh discretization and lets us transfer between the meshes without solving directly for a point-to-point correspondence. Our study has several limits: our experiments are limited only to LBS, while theoretical, there is no limit to other animation settings (e.g., deformation cages). Also, while we stressed our method on some challenging cases (noisy, broken, non-isometric shapes), we think many other cases would be interesting to analyze (e.g., topological issues, clutter). As future work, we plan to explore our new representation, for example, by studying its behave on different skeletons.

\section{Acknowledgements}

RM and SM are supported by the ERC Starting Grant No. 802554 (SPECGEO). This work is partially supported by the project of the Italian Ministry of Education, Universities and Research (MIUR) ”Dipartimenti di Eccellenza 2018-2022” of the Department of Computer Science of Sapienza University and the Department of Computer Science of Verona University

\bibliographystyle{ACM-Reference-Format}
\bibliography{acmart}










\end{document}